\documentclass[draftcls,11pt,onecolumn]{IEEEtran}
\usepackage{epsfig}
\usepackage{amsmath,amssymb}
\usepackage{xcolor}
\usepackage{algorithm}
\usepackage{algorithmic}
\usepackage{multirow}
\usepackage{graphicx}
\usepackage{stfloats}
\usepackage{amsfonts}
\usepackage{mathrsfs}
\usepackage[noblocks]{authblk}
\usepackage[compress]{cite}
\usepackage{bm}
\usepackage{changebar}
\usepackage{longtable}
\usepackage{chngpage}
\usepackage{array}
\usepackage{booktabs}
\usepackage[printonlyused]{acronym}
\usepackage{geometry}

\begin{document}

\title{ \LARGE Optimal Transmit Beamforming for Secure SWIPT in Heterogeneous Networks}
\author{Bin Li, Zesong Fei,~\IEEEmembership{Senior Member,~IEEE}, and Zheng Chu,~\IEEEmembership{Member,~IEEE}
\thanks{This work was supported in part by the National Natural Science Foundation of China under Grant 61371075 and Grant 61421001, and in part by the 111 Project of China under Grant B14010. (Corresponding author: Zesong Fei)}

\thanks{B. Li and Z. Fei are with the School of Information and Electronics, Beijing Institute of Technology, Beijing 100081, China (e-mails: libin$\_$sun@bit.edu.cn; feizesong@bit.edu.cn).}
\thanks{Z. Chu is with the School of Science and Technology, Middlesex University, London,
NW4 4BT, U.K. (e-mail: z.chu@mdx.ac.uk).}
}
\maketitle

\begin{abstract}
This letter investigates the artificial noise aided beamforming design for secure simultaneous wireless information and power transfer (SWIPT) in a two-tier downlink heterogeneous network, where one femtocell is overlaid with one macrocell in co-channel deployment.
Each energy receiver (ER) in femtocell can be considered as a potential eavesdropper for messages intended for information receiver (IR).
Our objective is to maximize the secrecy rate at IR subject to the signal-to-interference-plus noise
ratio (SINR) requirements of macro users (MUs), transmit power constraint and energy harvesting constraint.
Due to the non-convexity of the formulated problem, it cannot be solved directly.
Thus, we propose a novel reformulation by using first-order
Taylor expansion and successive convex approximation (SCA) techniques.
Furthermore, an SCA-based algorithm with low complexity is proposed to arrive at provably convergent solution.
Finally, numerical results evaluate the performance of the proposed algorithm.
\end{abstract}


\section{Introduction \label{a}}
To provide higher data rate for $5$G wireless communications, heterogeneous network (HetNet) is emerging as a promising network densification architecture and has been hailed as a key solution \cite{Hossain2014Commun}.
In HetNets, smallcells (e.g., picocells and femtocells) are deployed within the coverage of a macrocell and are operated in the same spectrum,
the challenge is the resulting cross-tier interference.
On the other hand, simultaneous wireless information and power transfer (SWIPT) has been envisioned as an attractive technique for powering energy-constrained wireless networks \cite{XuJie2014TSP}.
Benefiting from the deployment of smallcells, the energy harvesting (EH) from serving base station (BS) is more efficient
due to the short access distance.
In the context of HetNets with SWIPT, related works were presented in \cite{Akbar2016TWC,Sheng2016JSAC}.

Since HetNet creates a multi-tier dynamic topology and thus its information security is critical.
Responding to this, physical layer security (PLS) \cite{Cumanan2016JSTSP}, as an alternative to traditional cryptographic techniques, has been introduced into the HetNets to realize secure communications \cite{Lv2015JSAC,Wang2016TCom,Wang2016Book}.
Facing SWIPT-enabled HetNets, information security is more critical due to the inherent openness of the multi-tier topology and the increased signal power for EH make information particularly vulnerable to be eavesdropped by the unsubscribed energy receivers (ERs) (ER has better channel than information receiver (IR)).
In addition, the complicated network architectures and the cross-tier interference make it much more challenging to realize PLS for SWIPT-enabled HetNets than that for SWIPT in a conventional single-tier cellular network.
To the best of our knowledge, by far little literature has investigated the secure beamforming in SWIPT-enabled HetNets.

In this letter, we consider a promising application of SWIPT to a two-tier HetNet (e.g., device-to-device
(D2D) and machine-to-machine (M2M)), where it consists of a macrocell with multiple MUs and a femtocell with one IR as well as multiple ERs. However, due to the broadcast nature of wireless channels, one critical issue arises that the messages sent to IR may be eavesdropped by ERs. To enhance secure transmission, the artificial noise (AN) is embedded at the intended signal to deteriorate the reception of ERs. Our goal is to maximize the secrecy rate at IR while satisfying the required constraints. Particularly, our main contributions is summarized as follows:
\begin{itemize}
\item In the SWIPT-enabled HetNet, femtocell base station (FBS) shares downlink
spectral resource with the macrocell base station (MBS), and the mutual interference between these two networks is taken into account. Based on this framework, we exploit co-channel interference as a useful resource to improve the secrecy rate of IR and energy harvesting at ERs.
\item With the non-convexity of the original optimization proplem, we propose a fresh perspective to reformulate it as a series of second-order cone (SOC) inequalities, which circumvents the rank-one constraint in the existing semidefinite relaxation (SDR) techniques. A successive convex approximation (SCA)-based iterative algorithm is proposed and achieve low-complexity.
\end{itemize}


\section{System Model \label{b}}

\begin{figure}
\setlength{\abovecaptionskip}{0.cm}
  \setlength{\belowcaptionskip}{-0.cm}
\centering
\includegraphics[width=2.9in]{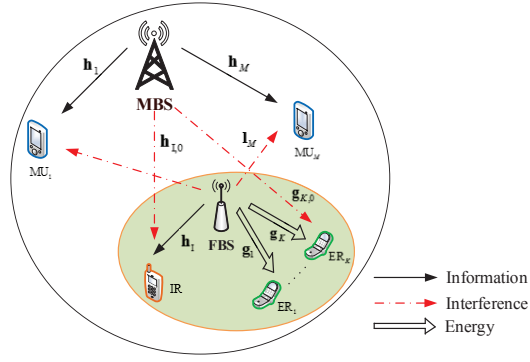}
\caption{System model of a two-tier HetNet supporting SWIPT.}
\label{sys}
\end{figure}

Consider a downlink two-tier HetNet where a FBS deploys with
a MBS, as shown in Fig. \ref{sys}.
The FBS serves $K+1$ femtocell users (FUs) and shares certain spectral resources as MBS serving $M$ MUs to improve the spectrum efficiency.
The MBS and FBS are equipped with $N_{\mathrm{M}} \geq M$ and $N_{\mathrm{F}} \geq K+1$ transmit antennas, respectively, whereas each MU and FU are equipped with a
single receive antenna. We assume that FBS is capable of performing wireless power transfer and exists two types of FUs in the femtocell,
i.e., one IR and $K$ ERs. Following the mechanism of \cite{LiuL2014TSP}, the separate IR and ER are adopted. Since the ERs may be malicious, they eavesdrop the information signal intended for IR. Thus, the ERs as
potential eavesdroppers should be taken into account.
Let $\mathcal{M} = \{1, \ldots, M\}$ denote the set of MUs and $\mathcal{K}= \{1, \ldots, K\}$ denote the set of ERs.
For notational simplicity, we assume that the $m$-th MU in the macrocell and the $k$-th ER in the femtocell are denoted by
MU$_m$ and ER$_k$, respectively.

The channel coefficients from MBS to MU$_m$, IR, and ER$_k$ are denoted by $\mathbf{h}_{m}\in \mathbb{C}^{N_\mathrm{M}\times 1}$, $\mathbf{h}_{\mathrm{I},0}\in \mathbb{C}^{N_\mathrm{M}\times 1}$, and $\mathbf{g}_{k,0}\in \mathbb{C}^{N_\mathrm{M}\times 1}$, respectively.
Likewise, the channel coefficients from FBS to
IR, ER$_k$, and MU$_m$ are denoted by $\mathbf{h}_{\mathrm{I}}\in \mathbb{C}^{N_\mathrm{F}\times 1}$,
$\mathbf{g}_{k}\in \mathbb{C}^{N_\mathrm{F}\times 1}$ and $\mathbf{l}_{m}\in \mathbb{C}^{N_\mathrm{F}\times 1}$, respectively.
All channels undergo flat-fading including large-scale fading, small-scale fading, and shadow fading.
Note that each ER is also a communication node in the same network which is assumed to be legitimate, active but do not have access to the information transmitted from MBS and FBS to IR. Full CSI of all receivers is assumed to be available at MBS and FBS. In practice, those CSI can be estimated via training and analog feedback \cite{Sheng2016JSAC,Cumanan2016JSTSP}.

To support secure communication and facilitate EH at ERs, an AN aided beamforming scheme is employed at
FBS. Therefore, the transmitted signal vector is denoted by
$\mathbf{x}=\mathbf{w}_\mathrm{I}s_\mathrm{I}+\mathbf{v}_\mathrm{E}$,
where $s_\mathrm{I}\sim \mathcal{CN}(0, 1)$ and $\mathbf{w}_\mathrm{I}$ denote the data symbol and beamforming vector, respectively. Hence, $\mathbf{w}_\mathrm{I}s_\mathrm{I}$ carries
the confidential information intended for IR.
$\mathbf{v}_\mathrm{E}\sim\mathcal{CN}(\mathbf{0},\mathbf{V}_\mathrm{E})$ denotes the energy-carrying AN vector invoked by
FBS, where $\mathbf{V}_\mathrm{E}$ represents the transmit covariance matrices of $\mathbf{v}_\mathrm{E}$.

Suppose $s_m\sim \mathcal{CN}(0, 1)$ is the data symbol transmitted by MBS intended for MU$_m$ and $\mathbf{w}_m$ is the
corresponding beamforming vector, the signal received at MU$_m$ by considering the co-channel interference can be expressed as
\begin{align}
  y_m=\mathbf{h}_m^H\mathbf{w}_ms_m+\sum_{i=1,i\neq m}^M\mathbf{h}_m^H\mathbf{w}_is_i+\mathbf{l}_m^H\mathbf{x}+n_m,
\end{align}
where $n_m\sim \mathcal{CN}(0, \sigma_m^2)$ denotes the additive white Gaussian noise (AWGN) at MU$_m$.
Then, the SINR at MU$_m$ can be represented as
\begin{align}
  \mathrm{SINR}_{m}=\frac{|\mathbf{h}_m^H\mathbf{w}_m|^2}{\sum_{i=1,i\neq m}^M|\mathbf{h}_m^H\mathbf{w}_{i}|^2+|\mathbf{l}_m^H\mathbf{w}_\mathrm{I}|^2
  +\mathbf{l}_m^H\mathbf{V}_\mathrm{E}\mathbf{l}_m+\sigma_m^2}
\end{align}

Since FBS desires to send the information-bearing signal $\mathbf{x}$ to IR while keeping it secret from ERs (suppose that it is an eavesdropper
to decode the message for IR instead of harvesting energy),
the signals received by both IR and ER$_k$ can be expressed, respectively, as
\begin{align}
  \mathrm{SINR}_{\mathrm{I}}=\frac{|\mathbf{h}_\mathrm{I}^H\mathbf{w}_\mathrm{I}|^2}
  {\sum_{m=1}^M|\mathbf{h}_{\mathrm{I},0}^H\mathbf{w}_{m}|^2+\mathbf{h}_\mathrm{I}^H\mathbf{V}_\mathrm{E}\mathbf{h}_\mathrm{I}+\sigma_\mathrm{I}^2},
  \label{SINR_I}
\end{align}
\begin{align}
  \mathrm{SINR}_{e,k}=\frac{|\mathbf{g}_k^H\mathbf{w}_\mathrm{I}|^2}
  {\sum_{m=1}^M|\mathbf{g}_{k,0}^H\mathbf{w}_{m}|^2+\mathbf{g}_k^H\mathbf{V}_\mathrm{E}\mathbf{g}_k+\sigma_{e,k}^2},
\label{SINR_Ek}
\end{align}
where $\sigma_\mathrm{I}^2$ and $\sigma_{e,k}^2$ denote the variances of AWGN at IR and ER$_k$, respectively.
As seen from (\ref{SINR_I}) and (\ref{SINR_Ek}), the IR and ERs suffer from the inter-tier interference in addition to intra-tier interference.
It is necessary to jointly design the beamformers at MBS and FBS in order to suppress the inter-cell interference
resulting from aggressive frequency reuse.

Then, the total transmit power of the whole network is given by
\begin{align}
P_{\rm tot}=\sum_{m=1}^M\|\mathbf{w}_\mathrm{m}\|^2+\|\mathbf{w}_\mathrm{I}\|^2+\mathrm{Tr}(\mathbf{V}_\mathrm{E})
\end{align}

On the other hand, based on the energy harvesting model \cite{LiuL2014TSP,Sheng2016JSAC}, the harvested energy at ER$_k$ is written as
\begin{align}
\mathrm{E}_k=\xi\left(|\mathbf{g}_{k}^H\mathbf{w}_\mathrm{I}|^2+\mathbf{g}_{k}^H\mathbf{V}_\mathrm{E}\mathbf{g}_{k}+\sigma_{e,k}^2\right)
\label{Ek}
\end{align}
where $\xi\in (0,1]$ is the energy conversion efficiency that accounts for
the loss converting the signal power to circuit power. It should be pointed out that the ER$_k$ is very difficult to harvest energy from MBS due to the long-distance transmission and poor wireless link.
Thus, the contribution of $\sum_{m=1}^M|\mathbf{g}_{k,0}^H\mathbf{w}_{m}|^2$ to EH is neglected here.

As such, the achievable instantaneous secrecy rate is formulated as
\begin{align}
R_{\mathrm{sec}}=\Big[\log_2(1+\mathrm{SINR}_\mathrm{I})-\max_{k\in\mathcal{K}} \log_2(1+\mathrm{SINR}_{e,k})\Big]^+
\end{align}
where the notation $[x]^+ = \max\{x,0\}$ is used.

Since FBS and MBS share the same radio resource, existing co-channel interference may degrade
the data rate of both the IR and ERs, as well as it is beneficial for ERs to harvest energy from the inter-tier interference.
Therefore, it is very nontrivial to properly designed secure beamforming and AN vectors to degrade the channels of ERs while having a minimal effect on IR.
Under such scenario, our objective is to seek the beamforming vectors $\mathbf{w}_m$, $\mathbf{w}_\mathrm{I}$, and AN covariance matrix
$\mathbf{V}_{\mathrm{E}}$ to maximize the secrecy rate at IR, while satisfying the SINR requirement for each MU, the total transmit power
and EH constraints. Hence, the optimization problem is formulated as
\begin{subequations}
\begin{align}
&\max_{\mathbf{w}_m,\mathbf{w}_\mathrm{I},\mathbf{v}_\mathrm{E}}~ R_{\mathrm{sec}}  \\
\mbox{s.t.}\quad
   \label{m1}
   &\mathrm{SINR}_m \geq \Gamma_m, ~\forall m\in\mathcal{M}, \\
   \label{m2}
   &P_{\mathrm{tot}} \leq P_{th}, \\
   \label{m3}
   &\mathrm{E}_k\geq Q_k,~\forall k\in\mathcal{K},
\end{align}
\label{eq1}
\end{subequations}

\noindent where $\Gamma_m$ is the prescribed target SINR of MU$_m$, $P_{th}$ is the maximum transmit power threshold and
$Q_k$ denotes the prescribed EH threshold at ER$_k$, respectively. It is clear that Problem (\ref{eq1}) is non-convex, since the objective
function constitutes a difference of two convex functions programming problem, which is hard to optimally solve due to prohibitively high computational complexity.
For practical purposes, we propose an SCA-based iterative algorithm to suboptimally solve Problem (\ref{eq1}).

\section{Secure Beamforming Design \label{c}}

In this section, we develop a suboptimal algorithm based on second-order cone programming (SOCP) relaxation for Problem (\ref{eq1}) to circumvent the rank-one constraint and achieve a low-complexity.
By introducing real-valued slack variables $\gamma$, $\gamma_\mathrm{I}$ and $\gamma_{\mathrm{E}}$, we rewrite Problem (\ref{eq1}) equivalently as
\begin{subequations}
\begin{align}
& \max_{\gamma, \gamma_\mathrm{I}, \gamma_{\mathrm{E}},\mathbf{w}_m,\mathbf{w}_\mathrm{I},\mathbf{v}_\mathrm{E}}~ \gamma  \\
\mbox{s.t.}\quad
   \label{e1}
   &\log_2(1+\gamma_\mathrm{I})-\log_2(1+\gamma_{\mathrm{E}})\geq \gamma, \\
   \label{e2}
   &\mathrm{SINR}_{\mathrm{I}}\geq \gamma_\mathrm{I}, \\
   \label{e3}
   &\mathrm{SINR}_{e,k}\leq \gamma_{\mathrm{E}}, \\
   \label{e4}
   &\mathrm{SINR}_m \geq \Gamma_m,  \\
   \label{e5}
   &P_{\mathrm{tot}} \leq P_{th}, \\
   \label{e6}
   &\mathrm{E}_k\geq Q_k,~\forall m\in\mathcal{M},~\forall k\in\mathcal{K}.
\end{align}
\label{SOC1}
\end{subequations}Note that Problem (\ref{SOC1}) is non-convex due to the coupled variables in constraints (\ref{e1})-(\ref{e3}) and (\ref{e6}).
To circumvent this predicament, we apply Taylor series expansion and SCA techniques \cite{Chu2016TVT} to reformulate the original problem as an SOCP.

\textit{Transformation of constraint (\ref{e2}):}
We first give the following lemma.

\textit{Lemma 1:} The hyperbolic constraint $z^2 \leq xy$ is equivalent to $\left\|[2z, x-y]^T\right\| \leq x+y$ when $x \geq 0$ and $y \geq 0$.

To make Problem (\ref{SOC1}) easier to tackle, we commence note that (\ref{e2}) can be equivalently transformed into
\begin{subequations}
\begin{align}
\label{g1}
   &|\mathbf{h}_\mathrm{I}^H\mathbf{w}_\mathrm{I}|^2\geq s_\mathrm{I}, \\
   \label{g2}
   &|\mathbf{h}_{\mathrm{I},0}^H\mathbf{w}_m|\leq s_m,~|\mathbf{h}_\mathrm{I}^H\mathbf{v}_\mathrm{E}|\leq s_\mathrm{E}, \\
   \label{g3}
   &\frac{s_\mathrm{I}}{\sum_{m=1}^Ms_m^2+s_\mathrm{E}^2+\sigma_\mathrm{I}^2}\geq \gamma_\mathrm{I},
\end{align}
\end{subequations}
where $s_\mathrm{I}$, $s_m$ and $s_\mathrm{E}$ are the introduced real-valued slack variables. Clearly, (\ref{g2}) is a convex constraint
and (\ref{g1}) is a non-convex constraint since the left side is a quadratic
function.
To deal with (\ref{g1}), we apply the first-order Taylor series expansion on $\tilde{\mathbf{w}}_\mathrm{I}$, we approximate the left side of (\ref{g1}) as
\begin{align}
|\mathbf{h}_\mathrm{I}^H\mathbf{w}_\mathrm{I}|^2=2\mathrm{Re}\{\tilde{\mathbf{w}}_\mathrm{I}^H\mathbf{H}_\mathrm{I}\mathbf{w}_\mathrm{I}\}-
\tilde{\mathbf{w}}_\mathrm{I}^H\mathbf{H}_\mathrm{I}\tilde{\mathbf{w}}_\mathrm{I}
\label{SOC2}
\end{align}

With (\ref{SOC2}), (\ref{g1}) becomes the following linear inequality
\begin{align}
2\mathrm{Re}\{\tilde{\mathbf{w}}_\mathrm{I}^H\mathbf{H}_\mathrm{I}\mathbf{w}_\mathrm{I}\}-
\tilde{\mathbf{w}}_\mathrm{I}^H\mathbf{H}_\mathrm{I}\tilde{\mathbf{w}}_\mathrm{I} \geq s_\mathrm{I}
\label{SOC3}
\end{align}

Now, we pay our attention to constraint (\ref{g3}).
By introducing real-valued slack variables $\mu_\mathrm{I}$ and $\eta_\mathrm{I}$, we further transform constraint (\ref{g3}) as
\begin{subequations}
\begin{align}
\label{h1}
   &s_\mathrm{I}\geq \mu_\mathrm{I}^2, ~\sum_{m=1}^M s_m^2+s_\mathrm{E}^2+\sigma_\mathrm{I}^2 \leq \eta_\mathrm{I}, \\
   \label{h2}
   &\frac{\mu_\mathrm{I}^2}{\eta_\mathrm{I}}\geq \gamma_\mathrm{I}.
\end{align}
\end{subequations}

By applying Lemma 1, (\ref{h1}) can be expressed as
\begin{align}
 \left\|[2\mu_\mathrm{I},s_\mathrm{I}-1]^T\right\|_2 \leq s_\mathrm{I}+1,
 \label{SOC4}
 \end{align}
\begin{align}
 \left\|[2s_1,\cdots,2s_M,2s_\mathrm{E},2\sigma_\mathrm{I},\eta_\mathrm{I}-1]^T \right\|_2 \leq \eta_\mathrm{I}+1.
\label{SOC41}
\end{align}

Recall that (\ref{h2}), we remark that the right side is an affine function and the left side is a quadratic-over-affine function.
According to the convexity of $\mu_\mathrm{I}^2/\eta_\mathrm{I}$,
applying first-order Taylor expansion on $(\tilde{\mu}_\mathrm{I}, \tilde{\eta}_\mathrm{I})$, we have \cite{Nguyen2015CL}
\begin{align}
2\left(\frac{\tilde{\mu}_\mathrm{I}}{\tilde{\eta}_\mathrm{I}}\right)\mu_\mathrm{I}-\left(\frac{\tilde{\mu}_\mathrm{I}}{\tilde{\eta}_\mathrm{I}}\right)^2\eta_\mathrm{I}\geq \gamma_\mathrm{I}
\label{SOC5}
\end{align}

\textit{Transformation of constraint (\ref{e3}):} In the same spirit, we split constraint (\ref{e3}) into
\begin{subequations}
\begin{align}
\label{o1}
   &|\mathbf{g}_k^H\mathbf{w}_\mathrm{I}|\leq t_k, \\
   \label{o2}
   &|\mathbf{g}_{k,0}^H\mathbf{w}_m|^2\geq t_{k,0},~|\mathbf{g}_{k}^H\mathbf{v}_\mathrm{E}|^2\geq t_{e,k}, \\
   \label{o3}
   &\frac{t_k^2}{\sum_{m=1}^Mt_{k,0}+t_{e,k}+\sigma_{e,k}^2}\leq \gamma_{\mathrm{E}},
\end{align}
\end{subequations}
where $t_k$, $t_{k,0}$, and $t_{e,k}$ are newly introduced real-valued slack variables.
Clearly, (\ref{o1}) is a convex constraint. Similar to (\ref{g1}), the two terms in (\ref{o2}) approximate as
\begin{align}
&2\mathrm{Re}\left\{\tilde{\mathbf{w}}_m^H\mathbf{G}_{k,0}\mathbf{w}_m\right\}-\tilde{\mathbf{w}}_m^H\mathbf{G}_{k,0}\tilde{\mathbf{w}}_m\geq t_{k,0}, \nonumber\\
&2\mathrm{Re}\left\{\tilde{\mathbf{v}}_\mathrm{E}^H\mathbf{G}_{k}\mathbf{v}_\mathrm{E}\right\}-\tilde{\mathbf{v}}_\mathrm{E}^H\mathbf{G}_{k}\tilde{\mathbf{v}}_\mathrm{E}\geq t_{e,k}.
\label{SOC6}
\end{align}

We can easily see that (\ref{o3}) is convex constraint. Let $b_k=\sum_{m=1}^Mt_{k,0}+t_{e,k}+\sigma_\mathrm{I}^2$,
according to Lemma 1, (\ref{o3}) can be equivalently transformed into the
following linear form
\begin{align}
 \left\|[2t_k,\gamma_{\mathrm{E}}-b_k]^T\right\|_2\leq \gamma_{\mathrm{E}}+b_k
\label{SOC7}
\end{align}

\textit{Transformation of constraint (\ref{e4}):}
We note that constraint (\ref{e4}) can be rewritten as
\begin{small}
\begin{align}
&\sum_{i=1,i\neq m}^M\mathbf{w}_i^H\mathbf{H}_m\mathbf{w}_i
   +(\mathbf{w}_\mathrm{I}+\mathbf{v}_\mathrm{E})^H\mathbf{L}_m(\mathbf{w}_\mathrm{I}+\mathbf{v}_\mathrm{E})+\sigma_m^2
\leq \frac{\mathbf{w}_m^H\mathbf{H}_m\mathbf{w}_m}{\Gamma_m}
\label{SOC8}
\end{align}
\end{small}

In light of \cite{Lv2015JSAC}, (\ref{SOC8}) can be transformed into the following SOC representation
\begin{align}
& \Big\|[2\mathbf{w}_1^H\mathbf{h}_m,\ldots,2\mathbf{w}_i^H\mathbf{h}_m,2\mathbf{w}_{i+1}^H\mathbf{h}_m,
 \ldots,2(\mathbf{w}_\mathrm{I}+\mathbf{v}_\mathrm{E})^H\mathbf{l}_m,2\sigma_m,
   \left(\mathrm{Re}(\mathbf{w}_m^H\mathbf{h}_m)/\sqrt{\Gamma_m}\right)-1]^T\Big\|_2   \nonumber\\
&\qquad  \qquad\qquad\qquad \qquad \qquad\qquad\qquad \qquad\qquad\qquad \leq \left(\mathrm{Re}(\mathbf{w}_m^H\mathbf{h}_m)/\sqrt{\Gamma_m}\right)+1, \nonumber\\
&\qquad  \qquad\qquad\qquad \qquad\qquad\qquad\qquad\qquad\qquad\qquad   \mathrm{Im}(\mathbf{w}_m^H\mathbf{h}_m)=0,~~\forall m\in\mathcal{M}.
\label{SOC9}
\end{align}


\textit{Transformation of constraint (\ref{e6}):} Next, we focus on non-convex constraint (\ref{e6}). Similarly, (\ref{e6}) can be transformed into convex form
\begin{align}
\xi[2\mathrm{Re}\{(\tilde{\mathbf{w}}_\mathrm{I}+\tilde{\mathbf{v}}_\mathrm{E})^H\mathbf{G}_k(\mathbf{w}_\mathrm{I}+\mathbf{v}_\mathrm{E})\}
-(\tilde{\mathbf{w}}_\mathrm{I}+\tilde{\mathbf{v}}_\mathrm{E})^H\mathbf{G}_k(\tilde{\mathbf{w}}_\mathrm{I}+\tilde{\mathbf{v}}_\mathrm{E})
+\sigma_{e,k}^2] \geq Q_k
\label{SOC10}
\end{align}

\textit{Transformation of constraint (\ref{e1}):}
Eventually, we return our attention to constraint (\ref{e1}). Since (\ref{e1}) is the difference of two concave functions, it is non-convex.
By introducing slack variable $c$ and performing first-order Taylor series expansion around point $\tilde{\gamma}_{\mathrm{E}}$,
(\ref{e1}) is rewritten as
\begin{align}
1+\gamma_\mathrm{I}\geq 2^c,  \label{SOC11}
\end{align}
\begin{align}
c-\log_2(1+\tilde{\gamma}_{\mathrm{E}})-\frac{{\gamma}_{\mathrm{E}}-\tilde{\gamma}_{\mathrm{E}}}{1+\tilde{\gamma}_{\mathrm{E}}}\geq \gamma.
\label{SOC12}
\end{align}

Although the exponential cone in (\ref{SOC11}) can be solved by existing nonlinear solvers (e.g. MOSEK and Fmincon), it requires more computation time in general.
To further reduce the computational complexity, according to a result in \cite{Nguyen2015CL}, (\ref{SOC11}) can be approximated in terms of a series of conic constraints as
\begin{align}
&\tau_0\leq 1+\gamma_\mathrm{I}, \nonumber\\
&\|[2+c/2^{q-1},1-\tau_1]\|_2\leq 1+\tau_1
, \nonumber\\
&\|[5/3+c/2^{q},1-\tau_2]\|_2\leq 1+\tau_2
, \nonumber\\
&\|[2\tau_1,1-\tau_3]\|_2\leq 1+\tau_3
, \nonumber\\
&\tau_2+\tau_3/24+19/72\leq \tau_4
, \nonumber\\
&\|[2\tau_{j-1},1-\tau_q]\|_2\leq 1+\tau_j, ~j\in \{5,\ldots,q+3\}\nonumber\\
&\|[2\tau_{q+3},1-\tau_0]\|_2\leq 1+\tau_0,
\label{SOC13}
\end{align}

\noindent where $\tau_j, \forall j=(0,1,\ldots,q+3)$, are the introduced slack variables,
and the accuracy of the approximation increases as $q$ increases.

Based on the above discussions, the approximated version of Problem (\ref{SOC1}) is given by
\begin{align}
&\max_{\substack{\mathbf{w}_m,\mathbf{w}_\mathrm{I},\mathbf{v}_\mathrm{E},\gamma,\gamma_\mathrm{I},\gamma_{\mathrm{E}},s_\mathrm{I}\\s_m,s_\mathrm{E}, \mu_\mathrm{I}, \eta_\mathrm{I}, t_k, t_{k,0},t_{e,k},c,\tau_j}}
\quad  \gamma_\mathrm{I}-\gamma_{\mathrm{E}} \nonumber\\
\mbox{s.t.}\quad
   &\mathrm{(\ref{e5})},\mathrm{(\ref{g2})},\mathrm{(\ref{SOC3})},\mathrm{(\ref{SOC4})},\mathrm{(\ref{SOC41})},\mathrm{(\ref{SOC5})},\mathrm{(\ref{o1})},\nonumber\\
   &\mathrm{(\ref{SOC6})},\mathrm{(\ref{SOC7})},\mathrm{(\ref{SOC9})},\mathrm{(\ref{SOC10})},\mathrm{(\ref{SOC12})},~\mathrm{and} ~\mathrm{(\ref{SOC13})}.
   \label{SOC14}
\end{align}

Note that Problem (\ref{SOC14}) is a convex SOCP problem, which can be efficiently solved by using
existing solvers, e.g., CVX \cite{Grant2010CVX}. The detailed procedure is summarized in Algorithm $1$.
\begin{algorithm}
\caption{Iterative Algorithm for Solving Problem (\ref{SOC14})}

\noindent \hangafter=1 \setlength{\hangindent}{4em}
\textbf{Input:~~}Set $n=0$, initialize $\tilde{\mathbf{w}}_m^{(0)}$, $\tilde{\mathbf{w}}_\mathrm{I}^{(0)}$, $\tilde{\mathbf{v}}_\mathrm{E}^{(0)}$, $\tilde{\mu}_\mathrm{I}^{(0)}$, $\tilde{\eta}_\mathrm{I}^{(0)}$, $\tilde{\gamma}_{\mathrm{E}}^{(0)}$ as the values which are feasible to
Problem (\ref{SOC14}).

\noindent \hangafter=1 \setlength{\hangindent}{4em}
\textbf{Step~1:~~}Solve the convex Problem (\ref{SOC14}) with ($\tilde{\mathbf{w}}_m^{(n)}$, $\tilde{\mathbf{w}}_\mathrm{I}^{(n)}$, $\tilde{\mathbf{v}}_\mathrm{E}^{(n)}$, $\tilde{\mu}_\mathrm{I}^{(n)}$, $\tilde{\eta}_\mathrm{I}^{(n)}$, $\tilde{\gamma}_{\mathrm{E}}^{(n)}$)
and obtain the optimal values (${\mathbf{w}_m^*}$, $\mathbf{w}_\mathrm{I}^*$, $\mathbf{v}_\mathrm{E}^*$, $\mu_\mathrm{I}^*$, $\eta_\mathrm{I}^*$, $\gamma_{\mathrm{E}}^*$).

\noindent \hangafter=1 \setlength{\hangindent}{4em}
\textbf{Step~2:~~}Update ($\tilde{\mathbf{w}}_m^{(n+1)}$, $\tilde{\mathbf{w}}_\mathrm{I}^{(n+1)}$, $\tilde{\mathbf{v}}_\mathrm{E}^{(n+1)}$, $\tilde{\mu}_\mathrm{I}^{(n+1)}$, $\tilde{\eta}_\mathrm{I}^{(n+1)}$, $\tilde{\gamma}_{\mathrm{E}}^{(n+1)}$)=(${\mathbf{w}_m^*}$, $\mathbf{w}_\mathrm{I}^*$, $\mathbf{v}_\mathrm{E}^*$, $\mu_\mathrm{I}^*$, $\eta_\mathrm{I}^*$, $\gamma_{\mathrm{E}}^*$).

~~~~~~~~~~~~$n=n+1$;

\noindent \hangafter=1 \setlength{\hangindent}{4em}
\textbf{Output:~~} $\mathbf{w}_m$ and $\mathbf{w}_\mathrm{I}$.

\end{algorithm}

\textit{Convergence Analysis:} From the above approximates, we can readily see that Problem (\ref{SOC14}) is convex, the optimal solutions can be obtained by solving (\ref{SOC14}) for a given ($\tilde{\mathbf{w}}_m$, $\tilde{\mathbf{w}}_\mathrm{I},\tilde{\mathbf{v}}_\mathrm{E},\tilde{\mu}_\mathrm{I},\tilde{\eta}_\mathrm{I},\tilde{\gamma}_{\mathrm{E}}$) in the $n$-th iteration.
Based on the update step in algorithm $1$ (i.e., Step $2$), the solutions in the $n$-th iteration are the feasible solutions in the ($n+1$)-th iteration.
This implies that the object value obtained in the ($n+1$)-th iteration is larger than or equal to that in the $n$-th iteration. In other
words, the secrecy rate in nondecreasing after each iteration. Furthermore, due to the power constraint, the secrecy rate is bounded. This conclusion illustrates the
convergence behavior of the proposed algorithm.

\textit{Complexity:}
According to \cite{Lv2015JSAC,Nguyen2015CL}, the main computational complexity for solving Problem (\ref{SOC14}) based on the proposed algorithm is $L_1\cdot\mathcal{O}\{N_\mathrm{M}M^{3.5}+N_\mathrm{M}^{3}M^{2.5}+N_\mathrm{F}(K+1)^{3.5}+N_\mathrm{F}^{3}(K+1)^{2.5}+(q+7)^3\}\log_2 (1/\epsilon)$, where $L_1$ is the number of iterations
and $\epsilon$ is the accuracy requirement.

\textit{Remark 1:} Our framework can be easily extended to the scenario consisting of
one macrocell and multiple femtocells, where the cooperation amongst multiple FBSs is necessary to enhance the secrecy
performance of the intended IR. The resultant optimization problem has to
consider different kinds of constraints with respect to the receivers and transmitters. The process of solving it is similar to the proposed
algorithm.


\section{Numerical results \label{d}}
In this section, numerical results are provided to show the performance of the proposed scheme. The parameters are set as $N_{\mathrm{M}} = 10$, $N_{\mathrm{F}} = 4$, $M=2$ and $K=2$, respectively.
In the considered propagation environment, the channel model adopted is given by \cite{Dahrouj2010TWC}
\begin{align}
\mathbf{h}_i=\sqrt{\beta(d_i)}\cdot\psi_i\cdot\varphi\cdot\tilde{\mathbf{h}}_i,~
\mathbf{g}_k=\sqrt{\beta(d_k)}\cdot\psi_k\cdot\varphi\cdot\tilde{\mathbf{g}}_k,
\end{align}
where $\beta(d)$ denotes the large scale fading coefficient given as $\beta(d)=10^{-\left(128.1+37.6\log_{10}(d)\right)/10}$ and $d$ represents the propagation distance.
$\psi_i$ and $\psi_k$ represent the log-normal shadow fading with zero mean and standard variance $8$dB. $\varphi$ is the transmit
antenna gain which is set to $15$dBi, and $\tilde{\mathbf{h}}_i$ and $\tilde{\mathbf{g}}_k$ are the multipath fading which are modeled as Rayleigh fading.
In simulations, the propagation distances from MBS to all receivers is $60$m, and
the distances from FBS to the MUs, IR and the ERs are $30$m, $20$m and $5$m, respectively.
For simplicity, we set the noise variances as $\sigma_m^2=\sigma_\mathrm{I}^2=\sigma_{e,k}^2=-100$dBm/Hz, the EH threshold as $Q_k=Q=15$dBm, the target SINR for each MU as $\Gamma_m=\Gamma=-10$dB, the energy conversion efficiency as $\xi=0.6$.

\begin{figure}
\centering
\includegraphics[width=2.8in]{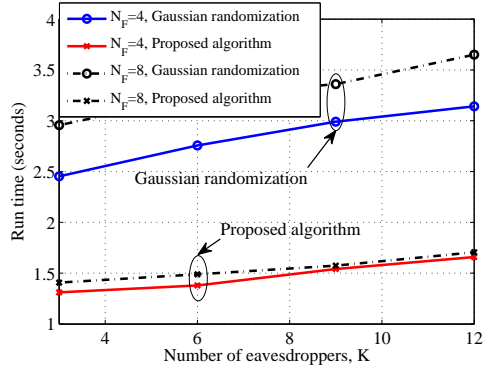}
\caption{Run time versus the number of eavesdroppers $K$ with $N_\mathrm{M}=10$.}
\label{fig:PF1}
\end{figure}

Fig. \ref{fig:PF1} compares the average run times of the proposed algorithm and SDP with rank relaxation scheme versus the number of eavesdroppers $K$ under $P_{th}=40$dBm.
It can be observed that the proposed algorithm is much faster than SDP via Gaussian randomization,
which implies that proposed algorithm has a lower computational complexity.
This is owing to the fact that the SOCP is well-structured convex form which can be solved more efficiently.

\begin{figure}[t]
\centering
\includegraphics[width=2.8in]{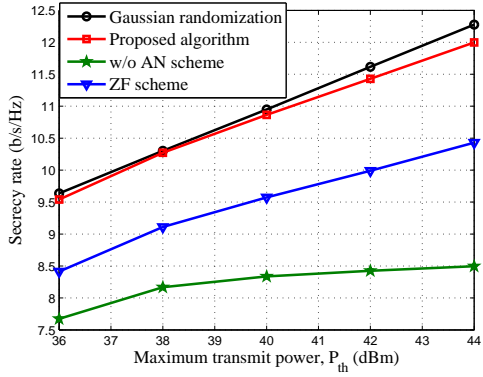}
\caption{Secrecy rate versus transmit power threshold $P_{th}$.}
\label{fig:PF3}
\end{figure}

For comparison, we consider three benchmark schemes, namely SDP via Gaussian
randomization scheme, without AN scheme (denoted as ``w/o AN scheme'') and zero-forcing beamforming scheme \cite{Sheng2016JSAC} (denoted as ``ZF scheme''). Note that the Gaussian randomization method represents the global optimal scheme and the ZF scheme represents no co-channel interference as well as the received signal is in the null space of eavesdropper.
As can be seen from Fig. \ref{fig:PF3}, the proposed algorithm outperforms the w/o AN scheme and ZF scheme, which validates the introduced co-channel interference is capable of improving the secrecy rate.
In low transmit power, the secrecy rate of the proposed algorithm is very close to that of the SDP via Gaussian
randomization. Furthermore, the performance gap is below 0.25b/s/Hz even in high transmit power region.

\section{Conclusion \label{e}}

In this letter, we investigated secure beamforming design in a two-tier HetNet with SWIPT.
Using SCA technique, we reformulated the secrecy rate maximization problem into a series of SOC forms. Then, an iterative algorithm with low-complexity was proposed to obtain the suboptimal solution.
Numerical results have been provided to corroborate the proposed scheme.

\bibliographystyle{IEEEtran}
\bibliography{refs}


\end{document}